# Three-Dimensional Anisotropic Thermal Conductivity Tensor of Single Crystalline $\beta$-Ga$_2$O$_3$

Puqing Jiang,[1,2] Xin Qian,[2] Xiaobo Li,[1,a] and Ronggui Yang[2,b]

[1]*School of Energy and Power Engineering, Huazhong University of Science and Technology, Wuhan, Hubei 430074, China*

[2]*Department of Mechanical Engineering, University of Colorado, Boulder, Colorado 80309, USA*

**Abstract:**

$\beta$-Ga$_2$O$_3$ has attracted considerable interest in recent years for high power electronics, where thermal properties of $\beta$-Ga$_2$O$_3$ play a critical role. The thermal conductivity of $\beta$-Ga$_2$O$_3$ is expected to be three-dimensionally (3D) anisotropic due to the monoclinic lattice structure. In this work, the 3D anisotropic thermal conductivity tensor of a (010)-oriented $\beta$-Ga$_2$O$_3$ single crystal was measured using a recently developed elliptical-beam time-domain thermoreflectance (TDTR) method. Thermal conductivity along any direction in the (010) plane as well as the one perpendicular to the (010) plane can be directly measured, from which the 3D directional distribution of the thermal conductivity can be derived. Our measured results suggest that at room temperature, the highest in-plane thermal conductivity is along a direction between [001] and [102], with a value of 13.3±1.8 W m$^{-1}$ K$^{-1}$, and the lowest in-plane thermal conductivity is close to the [100] direction, with a value of 9.5±1.8 W m$^{-1}$ K$^{-1}$. The through-plane thermal conductivity, which is along the [010] direction, has the highest value of 22.5±2.5 W m$^{-1}$ K$^{-1}$ among all the directions. Temperature-dependent thermal conductivity of $\beta$-Ga$_2$O$_3$ was also measured and compared with a theoretical model calculation to understand the temperature dependence and the role of impurity scattering.

---

[a] xbli35@hust.edu.cn
[b] Ronggui.Yang@Colorado.Edu

**Text:**

As a wide-bandgap semiconductor, β-Ga$_2$O$_3$ possesses many outstanding properties[1] such as incredibly large bandgap (4.5-4.9 eV),[2,3] high Baliga's figure of merit,[4,5] and excellent thermal stability. This enables Ga$_2$O$_3$ devices with an even higher breakdown voltage and efficiency than their SiC and GaN counterparts.[6] β-Ga$_2$O$_3$ has thus attracted a considerable interest in recent years for potential applications in the next-generation high-power and high-voltage devices.[7,8] Thermal properties of β-Ga$_2$O$_3$ can play a critical role in power electronic applications.

The thermal conductivity of β-Ga$_2$O$_3$ is expected to be 3D anisotropic due to the monoclinic lattice structure.[9] Thermal conductivities of β-Ga$_2$O$_3$ along some highly symmetric directions have been reported previously by several authors.[10-12] For example, Guo et al.[10] measured the through-plane thermal conductivities of several β-Ga$_2$O$_3$ samples with different orientations ([001], [100], [010], and [$\bar{2}$01]) using the time-domain thermoreflectance (TDTR) method from 80 to 495 K. They observed a ~$1/T$ dependence on the thermal conductivity for all the measured directions at temperatures >200 K and a $1/T^3$-$1/T^4$ dependence at temperatures <200 K. Handwerg et al.[11] reported the thermal conductivity of Mg-doped and undoped β-Ga$_2$O$_3$ bulk crystals along the [100] direction from 4 to 300 K measured using the 3ω method. Galazka et al.[12] also measured the thermal conductivity of β-Ga$_2$O$_3$ along the [010] direction using the laser flash method and reported a value of 21 W m$^{-1}$ K$^{-1}$ at room temperature. All the previous work reported the measured thermal conductivity along the through-plane directions of the samples, whereas the β-Ga$_2$O$_3$ crystals are usually only available with orientations along some highly symmetric crystallographic directions. Therefore, the 3D anisotropic thermal conductivity tensor of β-Ga$_2$O$_3$ has yet to be reported.

In this work, the 3D anisotropic thermal conductivity tensor of a β-Ga$_2$O$_3$ single crystal is measured using a recently developed elliptical-beam TDTR method.[13] By using a highly elliptical pump beam for heating the sample in TDTR experiments, the detected signals are selectively sensitive to the in-plane thermal conductivity along the short axis of the elliptical beam and the through-plane thermal conductivity but in different manners.[13] The 3D anisotropic thermal conductivity tensor of the sample can thus be derived from a series of measurements by rotating the elliptical pump beam. Our results suggest that at room temperature the highest in-plane thermal conductivity is along a direction between [001] and [102], with a value of 13.3±1.8 W m$^{-1}$ K$^{-1}$, and the lowest in-plane thermal conductivity is close to the [100] direction, with a value of 9.5±1.8 W m$^{-1}$ K$^{-1}$. The through-plane thermal conductivity, which is along the [010] direction, has the highest value of 22.5±2.5 W m$^{-1}$ K$^{-1}$ among all the directions. Temperature-dependent thermal conductivity of β-Ga$_2$O$_3$ was also measured, which agreed well with a theoretical model calculation. Comparison between the measurements and calculations help to elucidate the temperature dependence and the role of impurity scattering in thermal conductivity of β-Ga$_2$O$_3$.

Our sample is a (010)-oriented β-Ga$_2$O$_3$ single crystal purchased from Tamura Corporation (Japan).[14] This sample was unintentionally doped (mainly by silicon) with an impurity concentration of $\sim 1.3 \times 10^{17}$ cm$^{-3}$ measured by C-V profiling. Such a low concentration of impurity is expected to have a negligible effect on the lattice thermal conductivity and negligible electron contribution to the thermal conductivity.[15] The wafer has a dimension of $15 \times 10 \times 0.5$ mm, with the [010] direction along its through-plane and the [102] direction parallel to its breadth (see the illustration in Fig. 1(a)). A thin Al film with a nominal thickness of 100 nm was deposited on the (010) surface of the Ga$_2$O$_3$ substrate acting as a metal transducer for TDTR measurements. The actual film thickness was 105±4 nm determined by picosecond acoustics.[16]

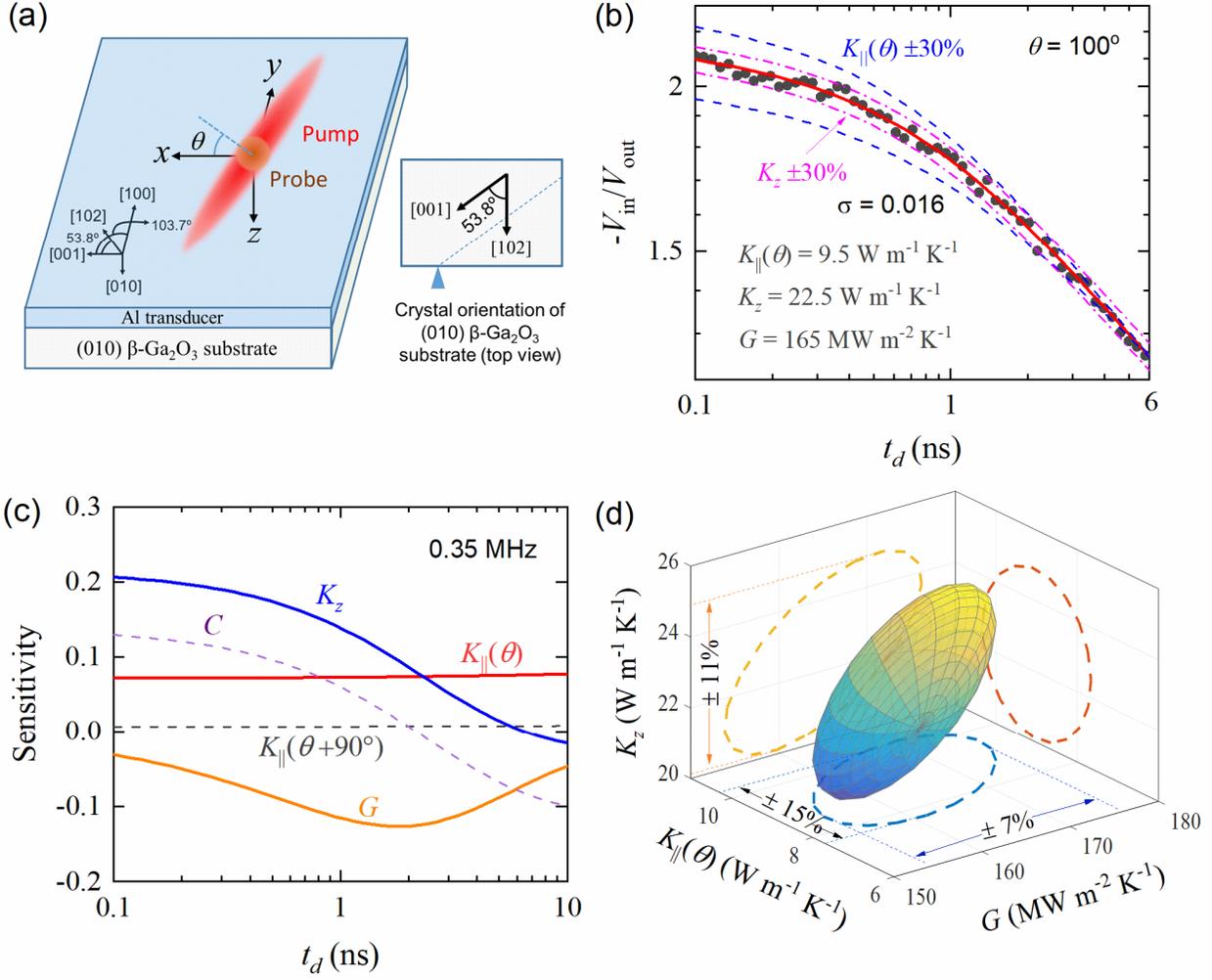

FIG. 1. (a) Illustration of sample configuration for the elliptical-beam TDTR measurements and the crystal orientation of (010) β-Ga$_2$O$_3$ substrate. (b) An example of the measured data (symbols) and model fitting (red solid lines) of $-V_{in}/V_{out}$ from 0.1 to 6 ns for the β-Ga$_2$O$_3$ sample with $\theta$=100°. 30% bounds of the best fitted $K_{||}(\theta)$ (dash-dot lines) and $K_z$ (dashed lines) are also included to show the signal sensitivity. The standard deviation $\sigma$ between the measured data points and the best fitted model calculation, which represents the signal noise level, is accounted for in the uncertainty estimation. (c) Sensitivity coefficients of thermal properties of the β-Ga$_2$O$_3$ substrate for the elliptical-beam TDTR measurements in (b). (d) Estimated uncertainties of $K_{||}(\theta)$, $K_z$ and $G$ when these three parameters are simultaneously determined from the measured data in (b).

Details of the elliptical-beam TDTR method and the TDTR system for thermal conductivity tensor measurements were described elsewhere.[13,17] To conduct the elliptical-beam TDTR measurements, a pair of cylindrical lenses were inserted in the pump path to generate a highly elliptical pump spot on the sample surface, whereas a circular beam was still used as the probe. To characterize the spot size, the pump spot was swept across the probe spot at a high modulation frequency of 10 MHz and a short positive delay time of 100 ps. The in-phase signal $V_{in}$ as a function of the offset distance $x_c$ was fitted by a Gaussian function $V_{in} \sim \exp(-x_c^2/w_0^2)$ to extract the root mean square average of the pump and probe $1/e^2$ radii along the sweeping direction.[18] The long and short radii of the elliptical laser spot were determined as 23 and 4.1 μm, respectively. A low modulation frequency of 0.35 MHz was used for all the thermal conductivity measurements. Thermal properties of the sample, including the in-plane thermal conductivity in the $\theta$ angle $K_{\parallel}(\theta)$, the through-plane thermal conductivity $K_z$, and the Al/Ga$_2$O$_3$ interface thermal conductance $G$, were determined by fitting the measured ratio signals $-V_{in}/V_{out}$ as a function of delay time using a thermal transport model.[13] Figure 1(b) shows an example of the measured signals compared with the model predictions. 30% bounds of the best fitted $K_z$ and $K_{\parallel}(\theta)$ values are also included as indications of the sensitivity of the signal to the thermal properties. It shows that the sensitivity to $K_z$ decreases dramatically as the delay time increases from 0.1 to 6 ns, whereas the sensitivity to $K_{\parallel}(\theta)$ remains constant over the whole delay time range. Since the signals are sensitive to $K_z$ and $K_{\parallel}(\theta)$ in different manners, these two parameters can be determined simultaneously by best-fitting the measured signals with the simulated signals. The sensitivity of the signal to each input parameter of the thermal transport model can be quantified by defining a sensitivity coefficient:[17]

$$S_\xi \equiv \frac{\partial \ln(-V_{in}/V_{out})}{\partial \ln \xi}, \tag{1}$$

where $\xi$ represents any input parameter in the thermal transport model. Figure 1(c) shows the sensitivity coefficients of the signal to different thermal properties of the substrate as a function of delay time. By using a highly elliptical pump beam for TDTR experiments, the detected signals are sensitive to $K_z$, $K_{\parallel}(\theta)$, and $G$ but all in different manners, meaning that these parameters are not strongly correlated and thus can be determined simultaneously. Note that the TDTR signals measured using an elliptical beam are essentially sensitive to the through-plane thermal effusivity ($e_z=\sqrt{K_zC}$) and the in-plane thermal diffusivity ($\alpha_{\parallel}(\theta)=K_{\parallel}(\theta)/C$). Since the heat capacity is isotropic, the sensitivity to the heat capacity of the substrate as shown in Fig. 1(c) is a combined effect of the sensitivities to $e_z$ and $\alpha_{\parallel}(\theta)$. To extract the thermal conductivities $K_z$ and $K_{\parallel}(\theta)$, we take the heat capacity of β-Ga$_2$O$_3$ from the literature[19] as a pre-known parameter, with an estimated uncertainty of <1%.[19] The uncertainties of $K_z$, $K_{\parallel}(\theta)$ and $G$ are then estimated using a multivariate error propagation formalism based on Jacobi matrices, which was first developed by Yang et al.[20] and widely used in some related works.[21-24] The effect of the signal noise, manifested as the standard deviation $\sigma$ between the measured data points and the thermal model prediction, is also included in the estimated uncertainty. Figure 1(d) shows that from a typical elliptical-beam TDTR measurement, $K_z$ and $K_{\parallel}(\theta)$ of β-Ga$_2$O$_3$ can usually be determined with an uncertainty of 11% and 15%, respectively.

The measured in-plane thermal conductivity tensor of the (010) β-Ga$_2$O$_3$ at room temperature is shown in Fig. 2(a). The $\theta$ angle is defined in the way with $\theta=0°$ parallel to the [001] direction, as illustrated in Fig. 1(a). The measured data are then automatically fitted using a rotated ellipse with both the rotation angle and the two axes of the ellipse as adjustable parameters. From Fig. 2(a), thermal conductivity along any direction in the (010) plane can be easily extracted by the intersection between the direction line and the ellipse of the thermal conductivity tensor. Specifically, some important crystallographic directions, including [100], [001], [102], [$\bar{2}$01], and ⊥[$\bar{2}$01], have been labeled out in Fig. 2(a). Here the notation ⊥[$\bar{2}$01] is used to denote the direction perpendicular to the [$\bar{2}$01] crystallographic direction. The results suggest that at room temperature, the highest in-plane thermal conductivity is along a direction between [001] and [102] and close to the ⊥[2̄01] direction, with a value of 13.3±1.8 W m⁻¹ K⁻¹, and the lowest in-plane thermal conductivity is close to the [100] direction, with a value of 9.5±1.8 W m⁻¹ K⁻¹. Each data point shown in Fig. 2(a) corresponds to one independent measurement of the through-plane thermal conductivity, which is along the [010] direction. Excellent consistency was achieved for the independently measured through-plane thermal conductivities, varying in the range 22-23 W m⁻¹ K⁻¹. The measurements indicate a general trend that the thermal conductivity is larger along the crystal orientation of a smaller lattice constant due to the stronger interatomic force of the lattice.

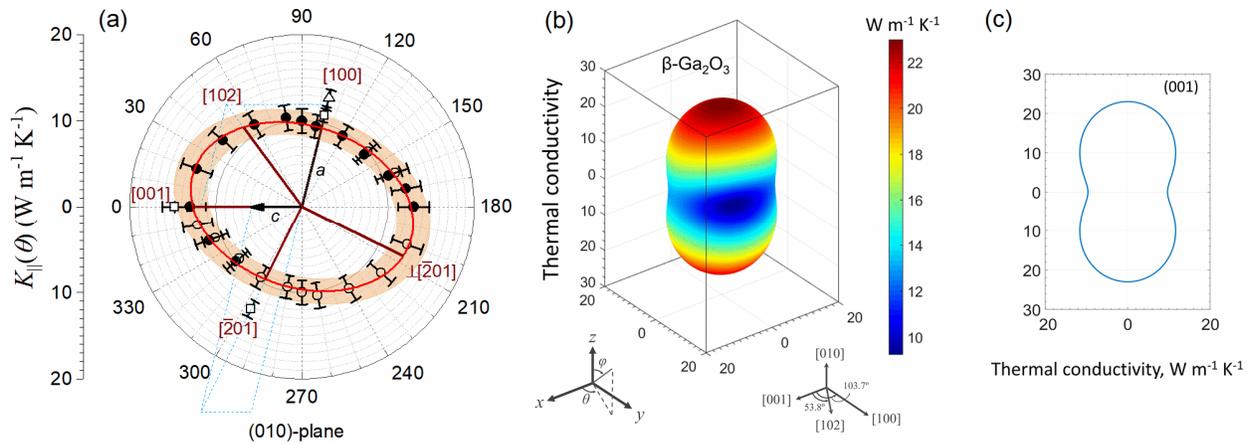

FIG. 2. (Color online) (a) In-plane thermal conductivity tensor of the (010)-oriented β-Ga₂O₃ substrate at room temperature measured using the elliptical-beam TDTR method. The solid symbols are the independent measurements and the open circles are their 180º reflections. The red curve is the best fit of the data points using a rotated ellipse. Some highly symmetric directions are labeled out in this plot. Some literature data including those from Guo et al.[10] (open squares) and Handwerg et al.[11] (open triangles) are also included for comparison. (b) 3D directional thermal conductivity of β-Ga₂O₃ from the current measurements at room temperature. (3) Projection of the 3D thermal conductivity tensor of β-Ga₂O₃ on the (001) plane.

From the measured in-plane thermal conductivity tensor and the through-plane thermal conductivity, we can calculate the full thermal conductivity tensor and the 3D directional thermal conductivity in a way discussed below. Since β-Ga₂O₃ is a monoclinic crystal, the thermal conductivity tensor takes the form of

$$\mathbf{K} = \begin{bmatrix} K_{xx} & K_{xy} & 0 \\ K_{xx} & K_{yy} & 0 \\ 0 & 0 & K_{zz} \end{bmatrix}, \tag{2}$$

where the subscripts $x$, $y$ and $z$ correspond to the Cartesian axes defined in Fig. 1(a). The $\mathbf{K}$ tensor is symmetric due to the Onsager reciprocal relation,[25,26] and the components $K_{xz}$ and $K_{yz}$ are zero due to the monoclinic symmetry.[27] The through-plane thermal conductivity $K_{zz} = 22.5$ W m$^{-1}$ K$^{-1}$ has been determined from the measurements; thus only the in-plane thermal conductivity tensor $\begin{bmatrix} K_{xx} & K_{xy} \\ K_{xy} & K_{yy} \end{bmatrix}$ needs to be determined. The directional in-plane thermal conductivity is related to the in-plane thermal conductivity tensor as (See Supplementary Information Section S1 for detailed derivation):

$$K_\parallel(\theta) = K_{xx}\cos^2\theta + 2K_{xy}\sin\theta\cos\theta + K_{yy}\sin^2\theta, \tag{3}$$

where $\theta$ is the azimuthal angle. The components $K_{xx}$, $K_{xy}$ and $K_{yy}$ can thus be solved from $K_\parallel(\theta)$ at $\theta = 0°$, $45°$ and $90°$. Therefore, the full tensor of thermal conductivity is obtained as:

$$\mathbf{K} = \begin{bmatrix} 12.7 & -1.2 & 0 \\ -1.2 & 9.7 & 0 \\ 0 & 0 & 22.5 \end{bmatrix} \text{(W m}^{-1}\text{ K}^{-1}\text{).} \tag{4}$$

The minus sign of $K_{xy}$ here denotes that a temperature gradient along the $+y$ direction would result in a small deflection of heat flux along the $-x$ direction. After the full thermal conductivity tensor is determined, the 3D directional thermal conductivity $K(\theta, \varphi)$ can be expressed as:

$$K(\theta,\varphi) = K_\parallel(\theta)\sin^2\varphi + K_{zz}\cos^2\varphi, \tag{5}$$

where $\varphi$ is the polar angle. The 3D distribution of the directional thermal conductivity is shown in Fig. 2(b), and the projected thermal conductivity distribution in the (001) plane is shown in Fig.

2(c). Thermal conductivity of β-Ga$_2$O$_3$ along any arbitrary ($\theta, \varphi$) direction can be extracted from Fig. 2(b) by intersecting the direction line with the thermal conductivity surface.

We further measured thermal conductivity of β-Ga$_2$O$_3$ as a function of temperature. Figure 3 shows the temperature-dependent thermal conductivities of β-Ga$_2$O$_3$ along the [010] and [100] directions from the current measurements over the temperature range from 80 to 400 K (solid symbols), with the literature measurements[10-12] (open symbols) included for comparison. It is interesting to note that although the current measurements agree well with the measurements by Guo et al.[10] at high temperatures >200 K for both directions, the measurements by Guo et al.[10] at low temperatures <200 K are significantly higher than the current measurements. The current measurements exhibit a ~$1/T^{1.3}$ dependence throughout the experimental temperature range 80-400 K, whereas the measurements of Guo et al.[10] have the same temperature dependence at high temperatures >200 K but a $1/T^3$-$1/T^4$ dependence at low temperatures <200 K. Most evidently, the thermal conductivities of β-Ga$_2$O$_3$ measured by Guo et al.[10] at <100 K are 2-3 times higher than the current measurements, far beyond the measurement uncertainties. There could be several possible reasons for this discrepancy, see detailed discussions below.

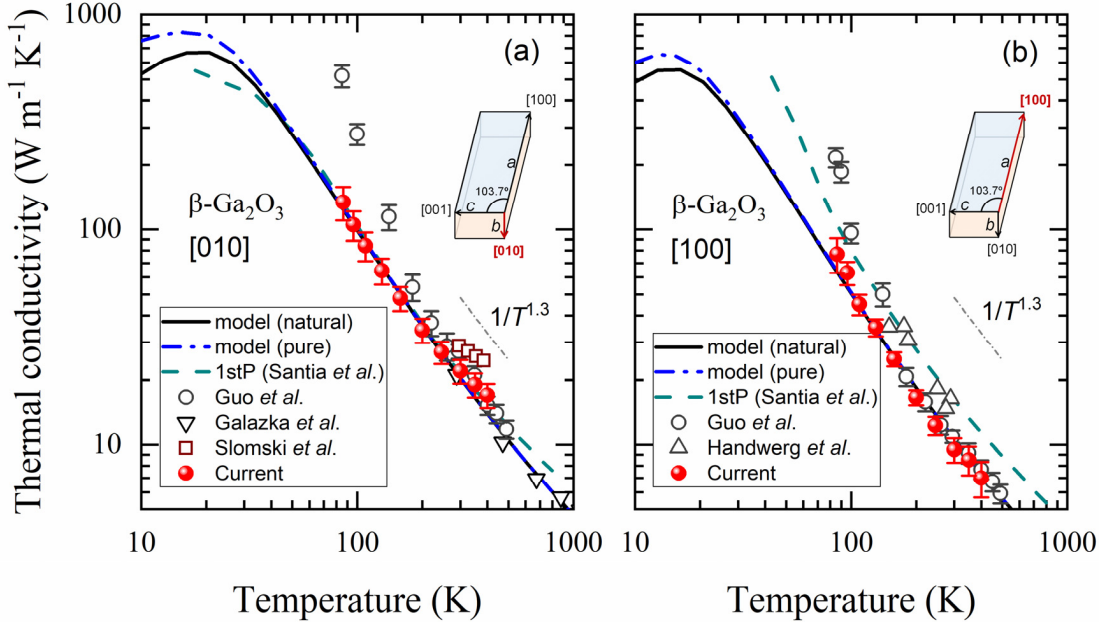

FIG. 3. Temperature-dependent thermal conductivity of β-Ga$_2$O$_3$ along the [010] and [100] directions from the current measurements (red solid symbols) are compared with the literature measurement[10-12,28] (different open symbols). Calculations using the modified Callaway model with pure (dash-dot lines) and natural isotope (solid lines) and the first-principles predictions from the literature[29] (dashed lines) are also included for comparison.

One possible reason could be that the samples might have different impurity concentrations, as impurity scattering usually dominantly affects the lattice thermal conductivity at low temperatures but not at high temperatures. To have a qualitative understanding of the temperature dependence and the effect of impurity scattering on the lattice thermal conductivity of β-Ga$_2$O$_3$, a theoretical model based on the Debye-Callaway formalism[30] was employed to calculate the temperature-dependent thermal conductivity of β-Ga$_2$O$_3$. This theoretical model was previously applied to study the isotope effect on the lattice thermal conductivity of a series of group IV and group III-V semiconductors, including wide-bandgap SiC and GaN.[30] In this theoretical model, a truncated linear dispersion is assumed, and the scattering rates due to phonon-phonon normal scattering, umklapp scattering and defect scattering are based on theories proposed by Herring,[31]

Peierls,[32] and Klemens.[33]. This model involves only two fitting parameters, *i.e.*, the Grüneisen parameters $\gamma_{L,T}$ for the longitudinal and transverse branches, which are assumed to be independent of temperature since the β-Ga$_2$O$_3$ crystal is very stable. In practice, we adjust the Grüneisen parameters until the calculated thermal conductivity agrees with the measurement at 300 K. We then study how the thermal conductivities of the samples with different impurity concentrations change as a function of temperature. By adjusting the Grüneisen parameters, the thermal conductivities at different temperatures change proportionally, which is because the temperature dependence terms and the impurity scattering terms in the model do not involve any Grüneisen parameter. The temperature dependence of thermal conductivity and the effect of impurity scattering are thus unaffected by the fitting parameters in the model. More details of the theoretical model can be found in Supplementary Information Section S2 and Ref. 30. The calculated results are plotted in Fig. 3 as the solid curves (naturally occuring) and dash-dot curves (isotopically pure) over a wide temperature range from 10 to 1000 K. The current measurements agree well with the calculations throughout the entire experimental temperature range. The results suggest that the phonon-isotope scattering should have a negligible effect at temperatures higher than 50 K. The unusually high thermal conductivities measured by Guo *et al*.[10] in the temperature range from 80 to 200 K cannot be explained by different impurity concentrations in the samples. The lattice thermal conductivity of β-Ga$_2$O$_3$ was also calculated by Santia *et al*.[29] using the first-principles method. Their results are also included as the dashed curves in Fig. 3. The first-principles calculated thermal conductivities agree well with the current measurements and calculations for the [010] direction, although being slightly higher than the current measurements and calculations for the [100] direction. We postulate that the first-principles calculations are free of unintended scattering sources (dislocations, stacking faults, impurities, etc.) that are likely to be present in the experimental samples and affect the measured thermal conductivities. The fact that our measured thermal conductivities are slightly lower than the first-principles calculations is a good indication on the validity of our measurements. The good agreement between the first-principles calculations and the current Callaway model calculation is also supportive of the temperature dependence of thermal conductivity as revealed from our measurements and calculations.

Another possible reason for the discrepancy could be the different heat capacity values of β-$Ga_2O_3$ used in these two studies when extracting the thermal conductivity values from TDTR experiments, especially for the large discrepancy at low temperatures < 200 K. Figure 4 summarizes the heat capacity values of β-$Ga_2O_3$ from several sources in the literature. For example, both King[19] and Adams and Johnston[34] have measured the heat capacity of β-$Ga_2O_3$ from room temperature down to 50 K, and their data differ by < 1.3% from each other. King[19] argued that their difference is because Adams and Johnston's sample contained over 1% impurities. On the other hand, Guo et al.[10] and Galazka et al.[12] used commercial differential scanning calorimeters (DSC) for their measurements and were only able to measure at relatively elevated temperatures > 123 K. To obtain the heat capacity values at low temperatures, Guo et al.[10] fitted their measured data to a Debye model, shown as the solid curve in Fig. 4. Figure 4 shows that data from the different sources agree well with each other at high temperatures > 200 K, but the fitted data by Guo et al.[10] deviate significantly from the literature measurements[19,34] at low temperatures <200 K. In particular, the heat capacity of β-$Ga_2O_3$ from Guo et al.[10] at 80 K is less than half the value of King[19]. The reason for the discrepancy can be that the Debye model used by Guo et al.[10] assumed a constant Debye temperature, which is inaccurate at intermediate temperatures, considering that the Debye temperature is temperature-dependent in a strict sense for most materials.[35] Since the TDTR experiments by Guo et al.[10] essentially measured the through-plane thermal effusivity ($\sqrt{K_z C}$) of the substrate, they would have thus obtained a much higher thermal conductivity value when a much lower heat capacity value was used.

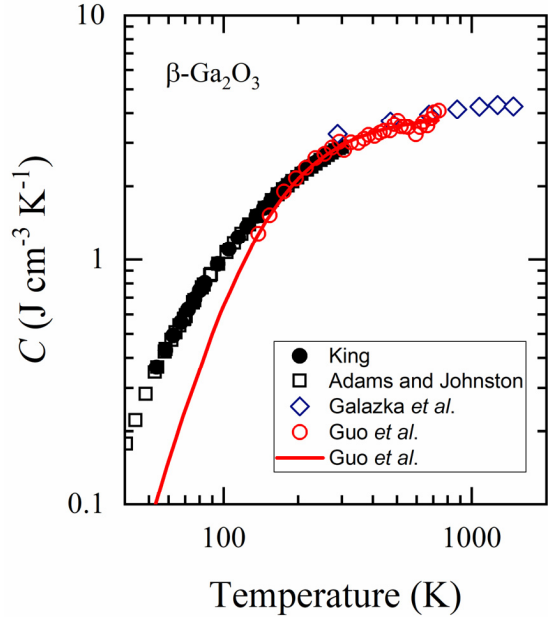

FIG. 4. Temperature-dependent volumetric heat capacity of β-Ga$_2$O$_3$ from the literature. The data by King[19] (51-298 K, solid symbols) and the one by Adams and Johnston[34] (15-300 K, open squares) were measured using adiabatic calorimetry. The data by Galazka et al.[12] (300-1470 K, open diamonds) were measured using DSC 404 from NETZSCH. The data by Guo et al.[10] (123-748 K, open circles) were measured using DSC-1 from Mettler Toledo. Note that the original data in the literature were presented with a unit of [J mol$^{-1}$ K$^{-1}$] or [J g$^{-1}$ K$^{-1}$]. A molar mass of 187.44 g mol$^{-1}$ and a density of 5.961 g cm$^{-3}$ were assumed[36] to convert the literature data into the volumetric heat capacity with a unit of [J cm$^{-3}$ K$^{-1}$].

In summary, the 3D anisotropic thermal conductivity tensor of a (010) β-Ga$_2$O$_3$ substrate was measured using an elliptical-beam TDTR method. Thermal conductivity along any direction in the (010) plane as well as the one perpendicular to the (010) plane can be directly measured, from which the full 3D thermal conductivity tensor can be derived. Among all the directions, the [010] direction has the highest thermal conductivity and the [100] direction has the lowest. The temperature-dependent thermal conductivity exhibits a $1/T^{1.3}$ dependence over the temperature range 80-400 K. A theoretical model calculation was conducted to elucidate the temperature dependence and the effect of impurity scattering on the lattice thermal conductivity of β-$Ga_2O_3$. The current measurement of temperature-dependent thermal conductivity agrees very well with our theoretical model calculations and a first-principles calculation from the literature.

See supplementary material for the derivation of thermal conductivity tensor and the theoretical model for lattice thermal conductivity of β-$Ga_2O_3$.

This work was supported by NSF Grant No. 1511195.

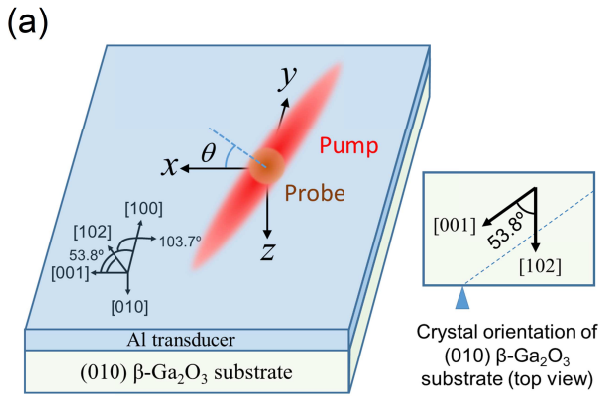
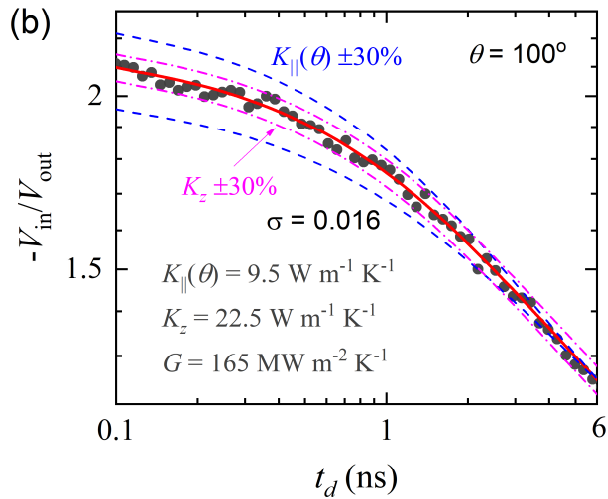
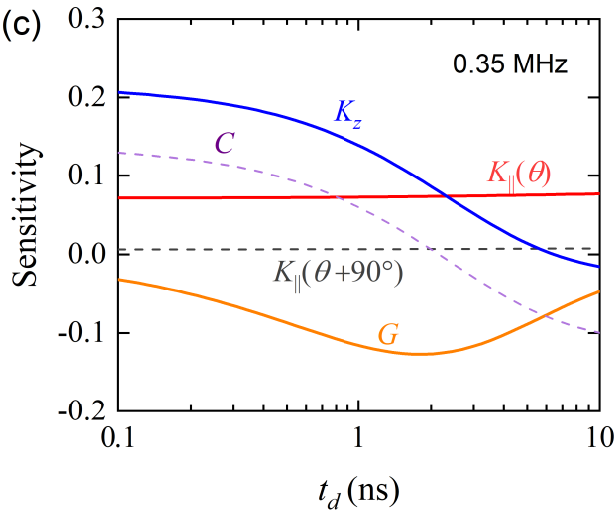
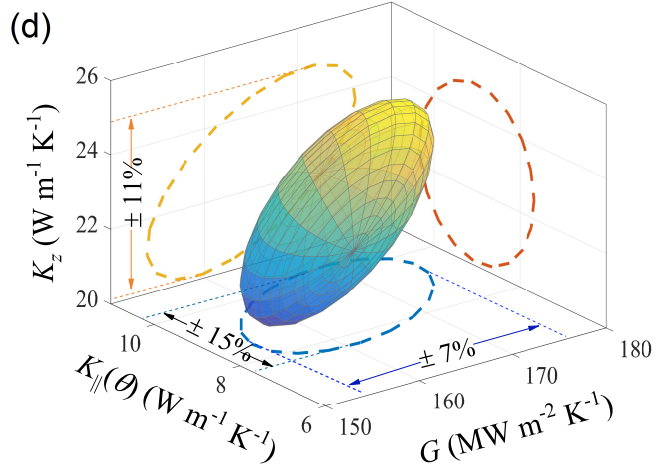

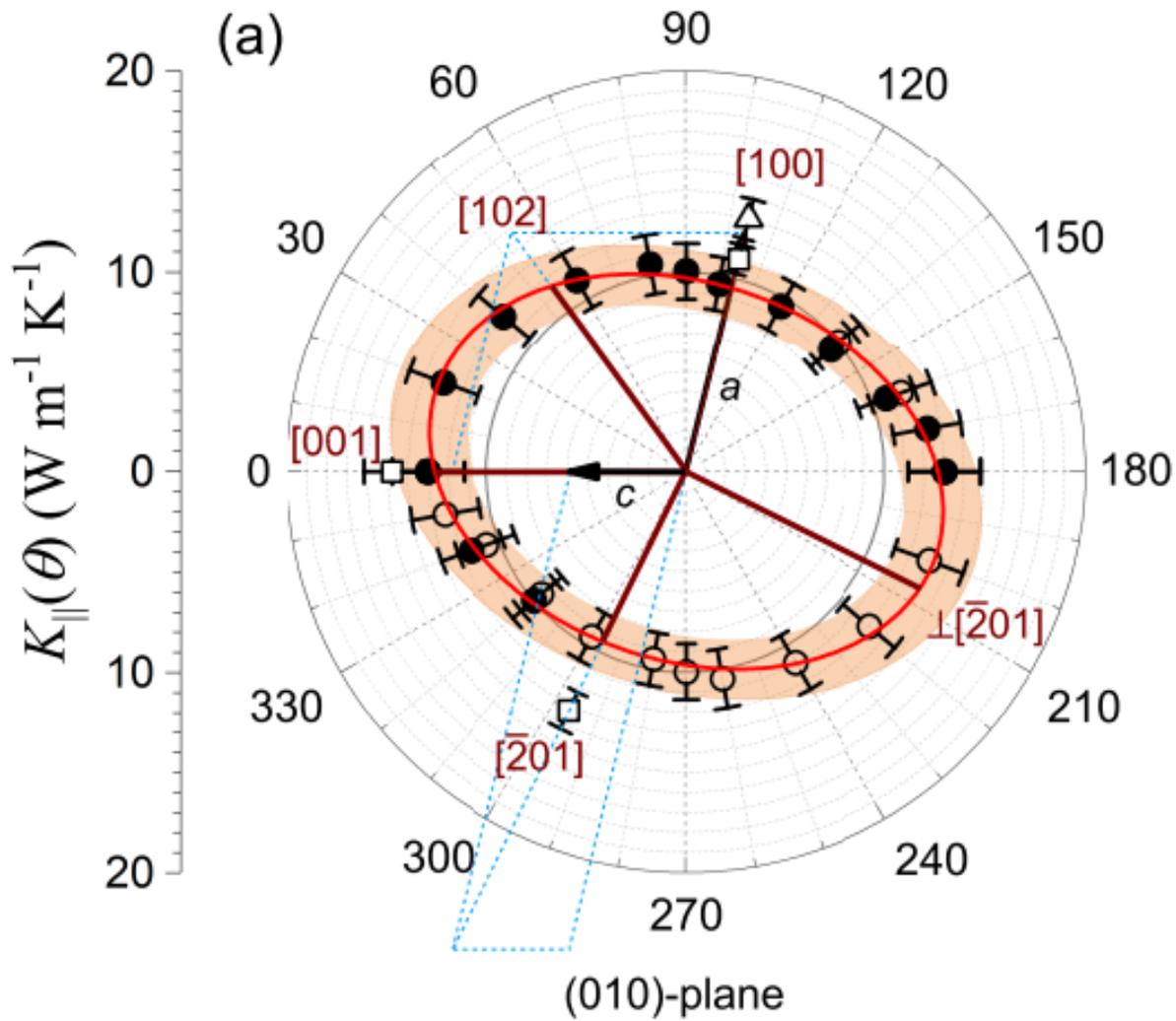
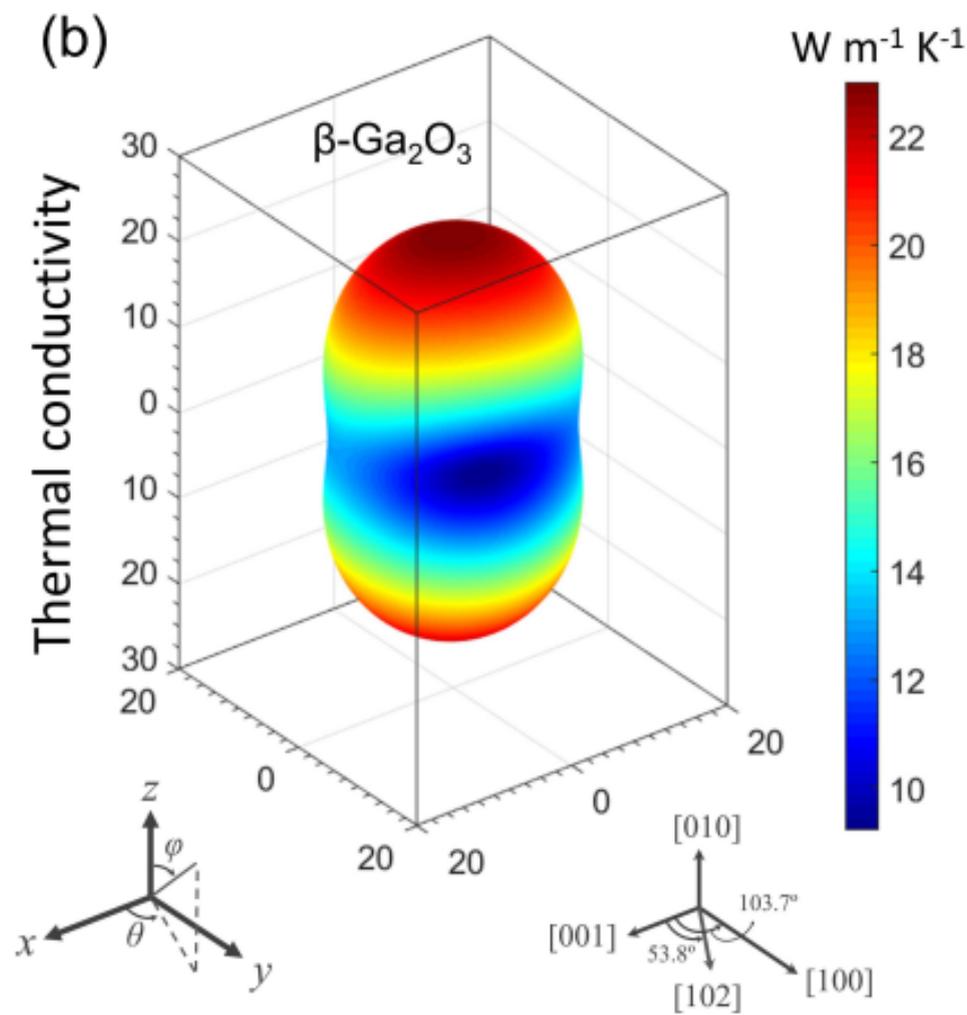
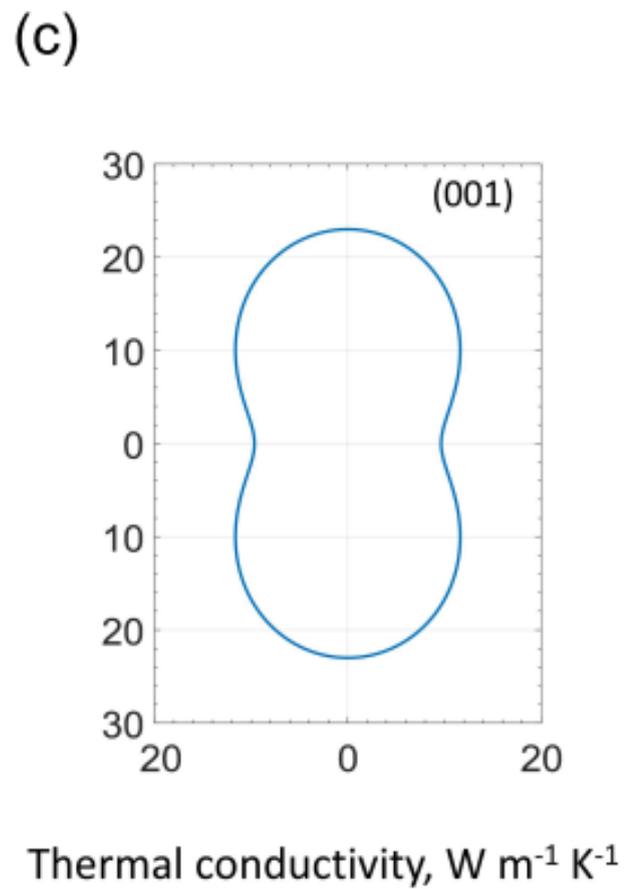

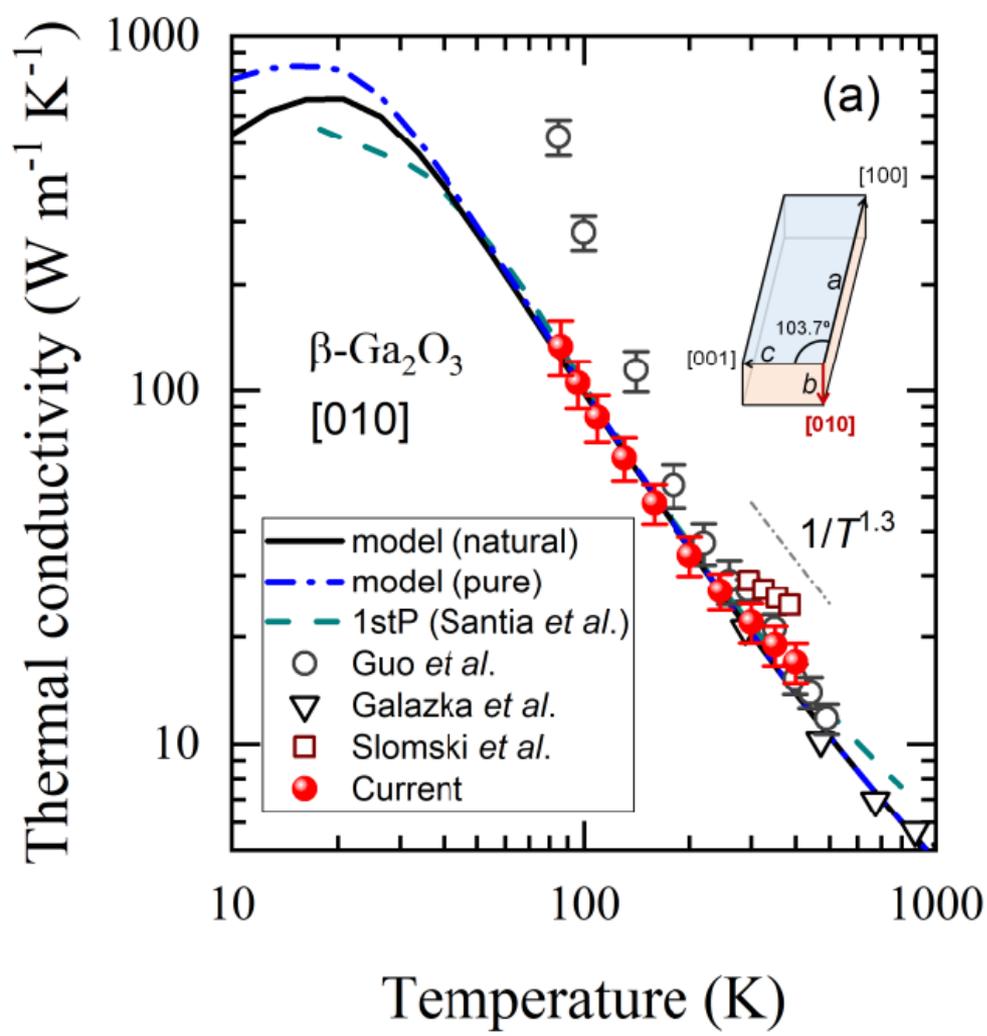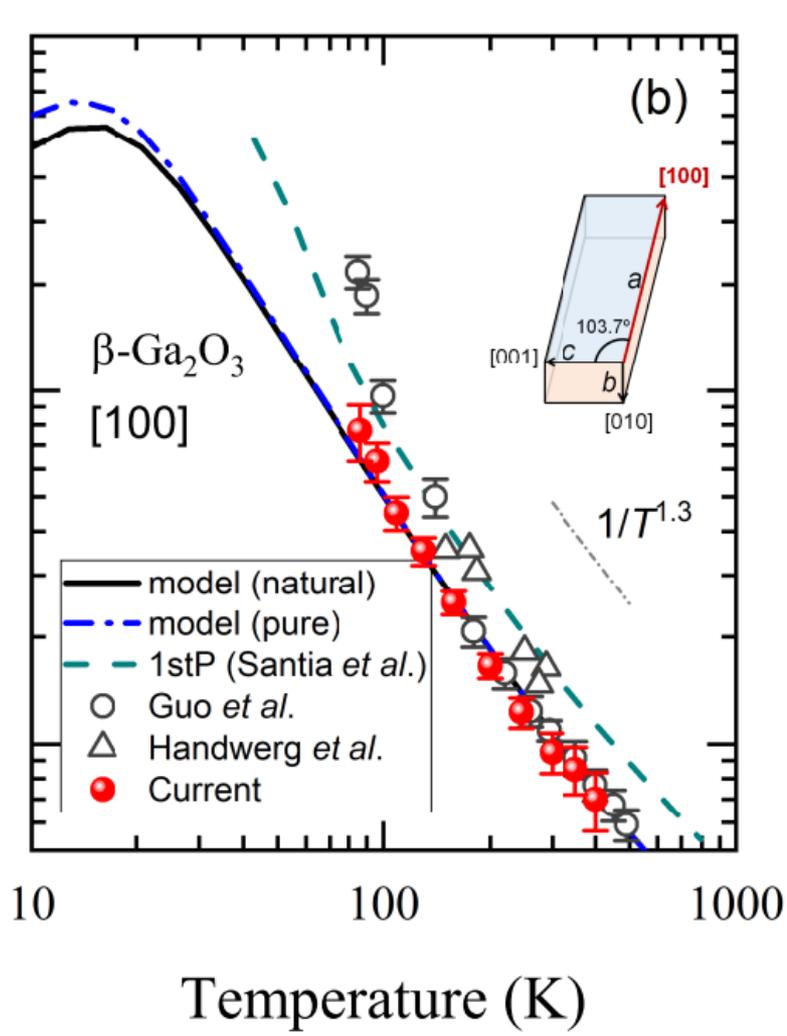

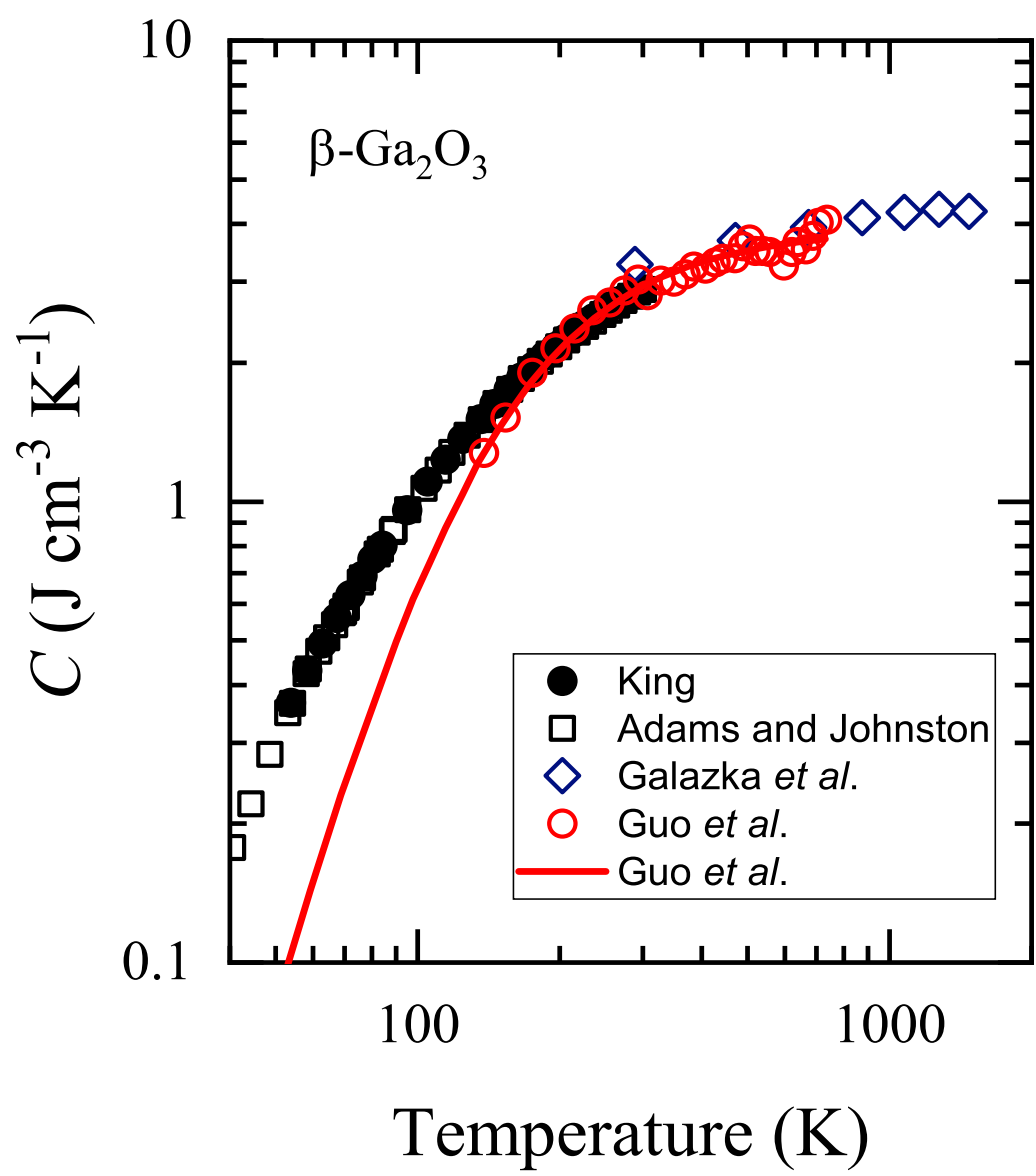

# Supplemental Information

# Three-Dimensional Anisotropic Thermal Conductivity Tensor of Single Crystalline β-Ga$_2$O$_3$

Puqing Jiang,[1,2] Xin Qian,[2] Xiaobo Li[1,a] and Ronggui Yang[2,b]

[1]*School of Energy and Power Engineering, Huazhong University of Science and Technology, Wuhan, Hubei 430074, China*

[2]*Department of Mechanical Engineering, University of Colorado, Boulder, Colorado 80309, USA*

## S1. Transformation of thermal conductivity tensor into directional thermal conductivity in a spherical coordinate system

Based on Fourier's law of heat conduction, the heat flux is related to the temperature gradient vector by the thermal conductivity tensor as:

$$\begin{bmatrix} q_x \\ q_y \\ q_z \end{bmatrix} = -\begin{bmatrix} K_{xx} & K_{xy} & 0 \\ K_{xy} & K_{yy} & 0 \\ 0 & 0 & K_{zz} \end{bmatrix} \begin{bmatrix} \dfrac{\partial T}{\partial x} \\ \dfrac{\partial T}{\partial y} \\ \dfrac{\partial T}{\partial z} \end{bmatrix}. \qquad (S1)$$

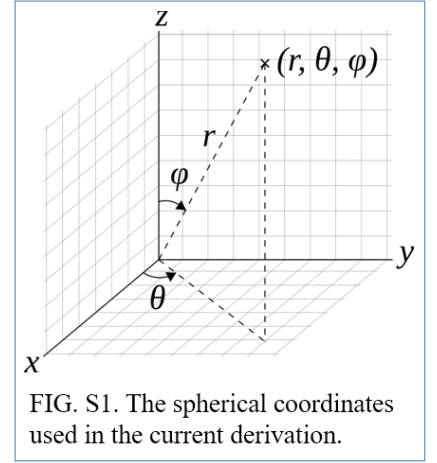

FIG. S1. The spherical coordinates used in the current derivation.

In the experiment, the temperature gradient in the radial direction $\partial T/\partial r$ with azimuthal angle $\theta$ and polar angle $\varphi$ is related to the temperature gradient vector as:

$$\begin{bmatrix} \dfrac{\partial T}{\partial x} \\ \dfrac{\partial T}{\partial y} \\ \dfrac{\partial T}{\partial z} \end{bmatrix} = \dfrac{\partial T}{\partial r} \begin{bmatrix} \sin\varphi\cos\theta \\ \sin\varphi\sin\theta \\ \cos\varphi \end{bmatrix}. \qquad (S2)$$

Similarly, for the heat flux in the radial direction with azimuthal angle $\theta$ and polar angle $\varphi$, we have:

---

[a] xbli35@hust.edu.cn
[b] Ronggui.Yang@Colorado.Edu

$$\begin{bmatrix} q_x \\ q_y \\ q_z \end{bmatrix} = q(\theta,\varphi) \begin{bmatrix} \sin\varphi\cos\theta \\ \sin\varphi\sin\theta \\ \cos\varphi \end{bmatrix}. \tag{S3}$$

Inverse matrix operation of Eq. (S3) yields:

$$q(\theta,\varphi) = \begin{bmatrix} \sin\varphi\cos\theta & \sin\varphi\sin\theta & \cos\varphi \end{bmatrix} \begin{bmatrix} q_x \\ q_y \\ q_z \end{bmatrix}. \tag{S4}$$

Combining Eqs. (S4), (S1) and (S2), we have:

$$\begin{aligned} q(\theta,\varphi) &= -\begin{bmatrix} \sin\varphi\cos\theta & \sin\varphi\sin\theta & \cos\varphi \end{bmatrix} \begin{bmatrix} K_{xx} & K_{xy} & 0 \\ K_{xy} & K_{yy} & 0 \\ 0 & 0 & K_{zz} \end{bmatrix} \frac{\partial T}{\partial r} \begin{bmatrix} \sin\varphi\cos\theta \\ \sin\varphi\sin\theta \\ \cos\varphi \end{bmatrix} \\ &= -\left[ K_{\parallel}(\theta)\sin^2\varphi + K_{zz}\cos^2\varphi \right] \frac{\partial T}{\partial r} = -K(\theta,\varphi)\frac{\partial T}{\partial r}. \end{aligned} \tag{S5}$$

We can thus write the 3D directional thermal conductivity as:

$$K(\theta,\varphi) = K_{\parallel}(\theta)\sin^2\varphi + K_{zz}\cos^2\varphi, \tag{S6}$$

where the in-plane thermal conductivity along the azimuthal angle $\theta$ is expressed as:

$$K_{\parallel}(\theta) = K_{xx}\cos^2\theta + 2K_{xy}\sin\theta\cos\theta + K_{yy}\sin^2\theta. \tag{S7}$$

## S2. The theoretical model based on Debye-Callaway formalism for lattice thermal conductivity of β-Ga$_2$O$_3$

In the theoretical model based on Debye-Callaway formalism,[1] the lattice thermal conductivity is calculated as a sum over one longitudinal ($K_L$) and two transverse ($K_{T_1}$ and $K_{T_2}$) phonon branches:

$$K = K_L + K_{T_1} + K_{T_2}$$

$$K_L = \frac{1}{3}\frac{k_B^4 T^3}{2\pi^2 \hbar^3 v_L}\left(I_{1,L} + I_{2,L}^2/I_{3,L}\right)$$

$$K_{T_1} = \frac{1}{3}\frac{k_B^4 T^3}{2\pi^2 \hbar^3 v_{T_1}}\left(I_{1,T_1} + I_{2,T_1}^2/I_{3,T_1}\right) \quad (S8)$$

$$K_{T_2} = \frac{1}{3}\frac{k_B^4 T^3}{2\pi^2 \hbar^3 v_{T_2}}\left(I_{1,T_2} + I_{2,T_2}^2/I_{3,T_2}\right)$$

where $k_B$ is the Boltzmann constant, $T$ is the temperature, $\hbar$ is the Planck constant, $v_L$, $v_{T_1}$ and $v_{T_2}$ are the speeds of sound for the longitudinal and transverse phonons, respectively. The integrals $I_{1,L}$, $I_{2,L}$ and $I_{3,L}$ for the longitudinal branch are defined below:

$$I_{1,L} = \int_0^{\theta_L/T} \tau_c^L \frac{x^4 e^x}{(e^x-1)^2} dx$$

$$I_{2,L} = \int_0^{\theta_L/T} \frac{\tau_c^L}{\tau_N^L} \frac{x^4 e^x}{(e^x-1)^2} dx \quad (S9)$$

$$I_{3,L} = \int_0^{\theta_L/T} \frac{\tau_c^L}{\tau_N^L \tau_R^L} \frac{x^4 e^x}{(e^x-1)^2} dx$$

where $x = \hbar\omega/k_B T$ is the normalized phonon frequency, $\omega$ is the phonon frequency, $\theta_L$ is the cutoff frequency for longitudinal phonons, $\tau_c$ is the combined phonon relaxation time defined as $\tau_c^{-1} = \tau_N^{-1} + \tau_R^{-1}$, $\tau_N$ is the relaxation time for the phonon-phonon normal scattering processes, and $\tau_R$ is the relaxation time for the resistive phonon scattering processes which includes the phonon-phonon umklapp scattering, phonon-impurity scattering, and phonon-boundary scattering. The corresponding integrals for the transverse branches $I_{1,T_1}$, $I_{2,T_1}$, $I_{3,T_1}$ and $I_{1,T_2}$, $I_{2,T_2}$, $I_{3,T_2}$ have the same expressions as in Eq. (S9), except that the cutoff frequency and relaxation times are respectively replaced by the values for the transverse branches.

Early theories by Herring,[2] Peierls,[3] and Klemens[4] have provided the theoretical expressions for the relaxation times of phonons due to different scattering processes, summarized as below.

a) *Phonon-phonon normal scattering*

Following the suggested form by Herring[2] and the approach of Asen-Palmer[5], the normal phonon scattering rates for longitudinal and transverse phonons are expressed as[1]

$$\left[\tau_N^L(x)\right]^{-1} = \frac{k_B^5 \gamma_L^2 V}{M \hbar^4 v_L^5} x^2 T^5$$

$$\left[\tau_N^T(x)\right]^{-1} = \frac{k_B^5 \gamma_T^2 V}{M \hbar^4 v_T^5} x T^5$$

(S10)

where $M$ (=6.2231 × 10$^{-26}$ kg/atom for β-Ga$_2$O$_3$) and $V$ (=1.0587 × 10$^{-29}$ m$^3$/atom for β-Ga$_2$O$_3$) are the average mass and volume of an atom in the crystal, $\gamma_L$ and $\gamma_T$ are the Grüneisen parameters for the longitudinal and transverse branches treated as adjustable parameters in the model. The values of sound velocities $v_L$, $v_{T_1}$ and $v_{T_2}$ for the [010] and [100] directions of β-Ga$_2$O$_3$ are given in Table S-1.

b) *Phonon-phonon umklapp scattering*

Following the suggested form by Peierls[3] and the approach of Slack and Galginaitis[6], the umklapp scattering rates have the following expressions:

$$\left[\tau_U^L(x)\right]^{-1} = \frac{k_B^2 \gamma_L^2}{M \hbar v_L^2 \theta_L} x^2 T^3 e^{-\theta_L/3T}$$

$$\left[\tau_U^T(x)\right]^{-1} = \frac{k_B^2 \gamma_T^2}{M \hbar v_T^2 \theta_T} x^2 T^3 e^{-\theta_T/3T}$$

(S11)

The values of cutoff frequencies $\theta_L$, $\theta_{T_1}$ and $\theta_{T_2}$ for the [010] and [100] directions of β-Ga$_2$O$_3$ are given in Table S-1.

c) *Rayleigh scattering due to impurities*

Following Klemens,[4] the Rayleigh scattering rates due to impurities in the crystals are expressed as:

$$\left[\tau_I^L(x)\right]^{-1} = \frac{V k_B^4 \Gamma}{4\pi \hbar^4 v_L^3} x^4 T^4$$

$$\left[\tau_I^T(x)\right]^{-1} = \frac{V k_B^4 \Gamma}{4\pi \hbar^4 v_T^3} x^4 T^4$$

(S12)

The mass-fluctuation phonon-scattering parameter $\Gamma$ for a binary compound $AB$ is given by[7]

$$\Gamma_{AB} = 2\left[\left(\frac{M_A}{M_A+M_B}\right)^2 \Gamma_A + \left(\frac{M_B}{M_A+M_B}\right)^2 \Gamma_B\right]$$

$$\Gamma_A = \sum_i c_i^A \left(\frac{m_i^A - M_A}{M_A}\right)^2 \qquad \text{(S13)}$$

$$\Gamma_B = \sum_i c_i^B \left(\frac{m_i^B - M_B}{M_B}\right)^2$$

where $M_A$ and $M_B$ are the average atomic mass of $A$ and $B$, respectively, $m_i^{A(B)}$ and $c_i^{A(B)}$ are the atomic mass and the atomic fraction of the $i$-th atom at site $A(B)$. Based on these formulas, we estimate a value of $\Gamma = 2.24 \times 10^{-4}$ for the phonon-isotope scattering rates of undoped β-Ga$_2$O$_3$ with a natural isotope composition. The natural isotope composition and the associated scattering parameter $\Gamma$ for Ga$_2$O$_3$ are listed in Table S-2.

To estimate the effect of the unintentionally doped Si, which has a concentration of ~$1.3 \times 10^{17}$ cm$^{-3}$, we treat the Ga$_2$O$_3$ compound as a monoatomic material[7] with an average atomic mass of $\bar{m}_{Ga_2O_3} = 37.4886$ u (unified atomic mass unit). The other information needed to calculate the scattering parameter $\Gamma$ due to the Si impurity includes: the atomic mass of Si, which is $\bar{m}_{Si} = 28.0855$ u, and the atomic density of β-Ga$_2$O$_3$, which is $9.1247 \times 10^{22}$ cm$^{-3}$. We thus estimate the atomic fraction of the Si dopant in our β-Ga$_2$O$_3$ substrate as $c_{Si} = \frac{1.3 \times 10^{17}}{9.1247 \times 10^{22}} = 1.42 \times 10^{-6}$, and the atomic fraction of Ga$_2$O$_3$ is $c_{Ga_2O_3} = 1 - c_{Si}$. The phonon-impurity scattering parameter $\Gamma$ due to the Si impurity is calculated to be $\Gamma_{Si} = 8.96 \times 10^{-8}$, which is four orders of magnitude lower than the phonon-isotope scattering parameter $\Gamma$. This indicates that the effect of Si impurity scattering is negligible compared to the mass isotope scattering. In our theoretical modeling, we compare the case of $\Gamma = 0$, which is isotopically pure, and $\Gamma = 2.24 \times 10^{-4}$, which is natural abundant, to examine the effect of phonon-isotope scattering on the thermal conductivity of β-Ga$_2$O$_3$.

d) *Phonon-boundary scattering*

The phonon-boundary scattering rate is assumed independent of temperature and frequency and is written as

$$\left(\tau_B^L\right)^{-1} = \frac{v_L}{d}$$
$$\left(\tau_B^T\right)^{-1} = \frac{v_T}{d}$$
(S14)

where $d$ is the thickness of the sample and is 0.5 mm for the current case.

The resistive scattering rate is the sum of scattering rates due to phonon-phonon umklapp scattering, Rayleigh scattering due to impurities, and scattering from the boundaries of the sample as:

$$\left(\tau_R^L\right)^{-1} = \left(\tau_U^L\right)^{-1} + \left(\tau_I^L\right)^{-1} + \left(\tau_B^L\right)^{-1}$$
$$\left(\tau_R^T\right)^{-1} = \left(\tau_U^T\right)^{-1} + \left(\tau_I^T\right)^{-1} + \left(\tau_B^T\right)^{-1}$$
(S15)

Fig. 3 in the main text presents the lattice thermal conductivity of β-Ga$_2$O$_3$ along the [010] and [100] directions calculated as a function of temperature using the model outlined above. It is worth mentioning that although the Callaway model involves fitting parameters $\gamma_{L,T}$, it still has the merit for revealing temperature dependence, because there are only two sources of temperature dependence in the Callaway model: the $N$-process scattering rate with $\tau_N^{-1} \propto T^3$ and the $U$-process scattering rate with $\tau_U^{-1} \propto T e^{-\theta/3T}$, the temperature dependence of both do not depend on the fitting parameters $\gamma_{L,T}$. We also note that the Callaway model has made the assumption of heat being carried only by acoustic phonons, despite the fact that optical phonons also contribute to the heat capacity and the overall thermal conductivity. Correspondingly, the model input parameters (listed in Table S-1) have been carefully chosen only to those acoustic modes to compensate the effect of the omission of optical phonons. For instance, the longitudinal and transverse Debye temperatures $\theta_L$, $\theta_{T1}$ and $\theta_{T2}$ were determined from the zone-boundary frequencies of those branches, respectively, rather than from the specific heat at low temperatures, which would average in all acoustic and optic modes and would be too high an estimate for the acoustic modes only. More detailed discussion can be found in Refs. 1 and 8. Meanwhile, the Gruneisen parameters for the acoustic branches are treated as fitting parameters, which also partially absorbs the error associated with the omission of optical phonons.

**Table S-1.** Zone-boundary frequencies $f_{L,T}$ and phonon velocities $v_{L,T}$ of longitudinal and transverse phonons for β-Ga$_2$O$_3$ along the [010] and [100] directions from the first-principles calculations in Ref. 9. $\theta_{L,T}$ are the Debye temperatures calculated from these cutoff frequencies following Ref. 1. $\gamma_{L,T}$ are the Grüneisen parameters obtained by fitting the calculated thermal conductivity at 300 K with the measurements.

| Direction | $f_L$ (THz) | $f_{T1}$ (THz) | $f_{T2}$ (THz) | $v_L$ (m s$^{-1}$) | $v_{T1}$ (m s$^{-1}$) | $v_{T2}$ (m s$^{-1}$) | $\theta_L$ (K) | $\theta_{T1}$ (K) | $\theta_{T2}$ (K) | $\gamma_L$ | $\gamma_T$ |
|---|---|---|---|---|---|---|---|---|---|---|---|
| [010] | 4.6 | 3.0 | 2.4 | 7270 | 3590 | 1960 | 220 | 144 | 115 | 1.1 | 0.85 |
| [100] | 4.0 | 2.6 | 2.2 | 6320 | 3840 | 2470 | 190 | 125 | 105 | 1.3 | 0.95 |

**Table S-2.** Natural isotope composition and associated scattering parameter Γ [Eq. (S13)] for Ga$_2$O$_3$.

| Material | Natural isotope composition | Γ (10$^{-4}$) | Material | Natural isotope composition | Γ (10$^{-4}$) |
|---|---|---|---|---|---|
| Ga | 60.1% $^{69}$Ga<br>39.9% $^{71}$Ga | 1.9847 | O | 99.76% $^{16}$O<br>0.04% $^{17}$O<br>0.2% $^{18}$O | 0.3285 |
| Ga$_2$O$_3$ | | 2.2399 | | | |